\documentclass{jps-cp}
\usepackage{txfonts} 
\usepackage{color}

\title{Wave-packet and entanglement dynamics in a non-Hermitian many-body system}

\author{Takahiro \textsc{Orito}$^{1}$ and Ken-Ichiro \textsc{Imura}$^{2}$}

\inst{$^{1}$Graduate School of Advanced Science and Engineering, Hiroshima University, Higashi-Hiroshima 739-8530, Japan \\
$^{2}$Institute of Industrial Science, The University of Tokyo, Kashiwa 277-8574, Japan}

\recdate{July 15, 2022}

\abst{
The static and dynamical properties of a one-dimensional quantum system described by
a non-Hermitian Hamiltonian of the so-called Hatano-Nelson type;
a tight-binding model with asymmetric (or non-reciprocal) hopping,
are studied.
The static properties of the model have been much studied;
the complex spectrum, skin effect and its topological interpretation, etc. 
Effects of disorder and inter-particle interaction,
especially, when they coexist, may be less understood.
Here, we will also focus on its dynamical properties
and reveal some unique features in the
wave-packet and entanglement (entropy) dynamics.
For that
some latest (original) results based on improved numerics 
(with this a system of larger system size $L$ becomes accessible)
are shown.
}

\kword{non-Hermitian, localization, topological, entanglement}

\begin{document}
\maketitle

\vspace{-1.5cm}
\section{Introduction}

Quantum mechanics described by a non-Hermitian Hamiltonian
attracts much attention recently, 
showing relevance, especially to low temperature physics.
An open quantum system is a typical system described by a non-Hermitian Hamiltonian.
Here,
we consider the 
Hatano-Nelson type
non-Hermitian Hamiltonian with asymmetric (or non-reciprocal) hopping [see, Eq. (\ref{ham})].
The original
Hatano-Nelson model
with onsite random (uncorrelated) potential disorder $W_j\in [-W/2,W/2]$
shows 
in spite of the one dimensionality of the model\cite{Anderson}
a localization-delocalization transition
at a finite critical disorder strength $W=W_c$
\cite{Hatano}.
Its spectrum under the periodic boundary condition (PBC) shows
a complex-real transition also at $W=W_c$.
Naturally, the existence of generally a finite imaginary part in its eigenvalues
is the hallmark of a non-Hermitian Hamiltonian.
The weak disorder regime of the Hatano-Nelson model under the PBC indeed 
falls on this category, while
under the open boundary condition (OBC) the spectrum becomes real.
Such a sensitivity to the boundary condition is another
peculiarity of a non-Hermitian Hamiltonian, already pointed out in the original work of
Hatano and Nelson.
The behavior of the corresponding eigen wave functions is also peculiar,
showing the so-called non-Hermitian skin effect
under the OBC.\cite{YaoWang}

Here, in this paper we report on
some specific consequences of such peculiar features of a non-Hermitian system
in its dynamics,
taking also into account
the effects of inter-particle interaction.
\vspace{-9mm}
\section{Model and its static properties} 
Let us consider the following variant of the
Hatano-Nelson model:
\begin{equation}
H=\sum_{j}[-(e^{g}\hat{c}_j^\dagger \hat{c}_{j+1}+e^{-g}\hat{c}_{j+1}^\dagger \hat{c}_{j})+V\hat{n}_{j}\hat{n}_{j+1}+W_j\hat{n}_j],
\label{ham}
\end{equation}
where 
$\hat{c}_j^\dagger$ ($\hat{c}_j$) creates (annihilates)
an electron at site $j$, 
and
$\hat{n}_{j}=\hat{c}_j^\dagger\hat{c}_j$.
$g$ is a measure of asymmetry in hopping,
while $V$ represents the strength of inter-particle (here, nearest-neighbor) interaction.
The last term represents onsite potential disorder at site $j$,
which here, we choose to be quasi-periodic; cf. Aubry-Andr\'e model\cite{AA}:
\begin{equation}
W_j=W\cos(2\pi\theta j+\theta_0),
\label{quasi}
\end{equation}
where $\theta$ is an irrational constant and chosen as $\theta=(\sqrt{5}-1)/2$,
and $\theta_0$ is an free parameter to take disorder average. 
Note that
for this type of (correlated) disorder,
localization transition occurs 
at $W=W_c=2e^g$
in the non-interacting case: $V=0$,
which 
(in contrast to the original Hatano-Nelson model with uncorrelated disorder)
remains finite even in the Hermitian limit: $g=0$.

\vspace{-0.5cm}
\begin{figure}[tbh]
\includegraphics[width=150mm]{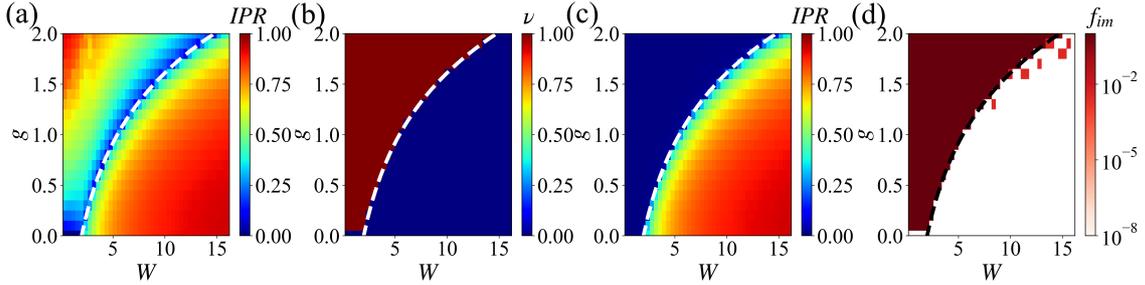}
\caption{A quartet of phase transitions (single-particle case): 
(a) IPR under OBC, (b) winding number $\nu$, (c) IPR under PBC
(d) Im ratio $f_{\rm im}$
in the space of parameters $(W,g)$.
}
\label{f1}
\end{figure}

\vspace{-0.5cm}
Under the PBC,
the model exhibits a complex-real transition of the eigenenergy at $W=W_c=2e^g$;
here,
we switch off the inter-particle interaction: $V=0$ for the moment.
Concomitantly,
the corresponding wave function shows a delocalization-localization transition
at the same value of $W=W_c$.
Under the OBC,
the eigen wave functions 
shows a peculiar damping, or an exponential decay
in the regime of weak disorder (Fig. 1 (a));
this is often called non-Hermitian skin effect.
Under the PBC, the same parameter region
corresponds to the regime of complex energy (Fig. 1 (d))
and delocalized wave function (Fig. 1 (c)).
To quantify these issues,
we have here considered the inverse participation ratio (IPR):
IPR $=\sum_j |\psi_j|^4$
as a measure of the localizability of the wave function $\psi_j$.
Note that
for a delocalized wave function IPR $\simeq 0$ 
(vanishes as $\simeq 1/L$, with $L$ being the size of the system), 
while 
for a localized wave function IPR $\simeq 1$; 
this includes the case of a wave function susceptible to the skin effect .
Fig. 1 (a) shows the IPR under OBC in the space of parameters $(W,g)$.
For a given $g$ one can observe that
IPR $\simeq 1$ (skin effect is effective)
in the regime of weak disorder, then
it once diminishes (IPR $\simeq 0$) in the critical regime $W\sim W_c$,
while in the regime of strong disorder: $W>W_c$
IPR increases again and takes a value $\simeq 1$.
Fig. 1 (c) shows the corresponding behavior of IPR under PBC, in which
IPR $\simeq 0$, i.e., the wave function is delocalized 
in the regime of weak disorder: $W<W_c$.
Fig. 1 (d) shows the variation of the ratio $f_{\rm im}=N_D/L$,
where $N_D$ is the number of the eigenenergy which has $Im(|\epsilon|)>10^{-13}$
in the same space of parameters $(W,g)$. 
In the delocalized phase, most of the eigenenergies are complex:
in contrast, in the localized phase, eigenenergies become real,
i.e., the delocalization-localization transition accompanies the complex-real transition. 

%
In Fig. 1 (b) 
a topological interpretation \cite{Gong}
is given to the skin effect.
For that
a winding number $\nu$ that counts how many times the (complex) spectrum under PBC
winds arounds the origin in the complex energy plane
as the crystal momentum $k$ once goes around the Brillouin zone.
In the presence of disorder (quasi-periodic potential),
or in case of the broken translational symmetry,
we introduce a flux $\Phi$ 
that twists the periodic boundary condition
and plays the role of $k$ in this case;
the winding number $\nu$ is then given as 
\begin{equation}
\nu=\int_0^{2\pi} {1\over {2\pi i}}d\Phi \partial_\Phi\log \det [H(\Phi)-E_0],
\label{winding}
\end{equation}
where $E_0$ is the base energy, which we choose 
here as $E_0=0$.
In the practical calculation, we also
employ a finite difference instead of the differential in Eq. \ref{winding}; 
$\nu$ is calculated by collecting 
finite differences of $H(\Phi)$ 
in $201$ points ($d\Phi={2\pi \over 201}$).
Fig. 1 (b)
the distribution of 
$\nu$
in the parameter space $(W,g)$
is shown and compared with the behavior of other indices: panels (a), (c), (d).
They all suggest that the transition occurs at $W=W_c=2e^g$.
\begin{figure}[tbh]
\includegraphics[width=150mm]{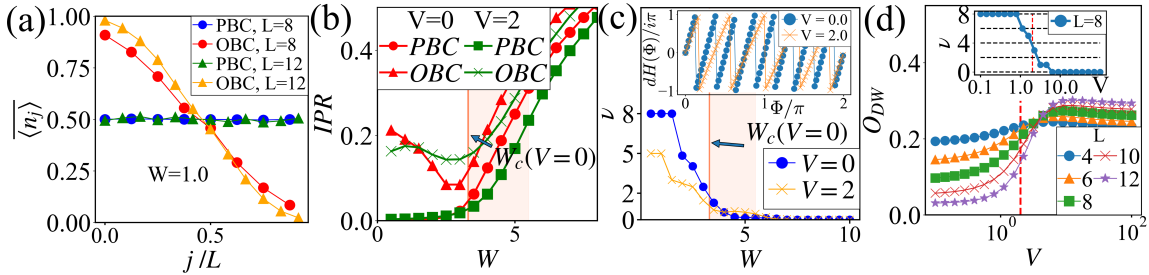}
\caption{Static properties (many-particle case): (a) local density $\langle n_j\rangle$ under OBC and PBC, 
(b) ${\cal IPR}$ as a function of the disorder strength $W$, 
(c) the winding number $\nu$ as a function of $W$.
50 disorder realizations. All eigenstates averaged.
Inset of the panel (c) shows $dH(\Phi)$ as a function of $\Phi$ with W=0.5 and $\theta_0=0$. 
The shaded region indicate the crossover regime;
$W_c$ is now renormalized by $V$.
(d) Charge density-wave order parameter $O_{DW}$ of ground state (under OBC) as a function $V$. The red dashed line locates $V=2$. Inset shows the winding number $\nu$ as a function of $V$ with W=0 and $\theta_0=0$.}
\label{f2}
\end{figure}

\vspace{-0.5cm}
In the case of a single particle we have seen so far 
all the eigenstates under the OBC
tend to be localized exponentially toward an end of the system,
and perhaps in an extreme case, localized to a single site at the boundary.
%
Four panels of Fig. 2 show that
the above single-particle scenario holds also true
in the many-particle case;
here we consider the case of $N=L/2$ particles (half-filling).
Fig. 2 (a) shows the distribution of the local density 
$\langle n_j\rangle=\langle \mu|n_j|\mu \rangle$
for $L=8, 12$,
where 
the label $\mu$ specifies a many-body eigenstate.
Under the OBC,
$\langle n_j\rangle$ exhibits 
an asymmetric density profile;
particles are predominantly located on the left half of the system.
In this ``many-body skin effect''\cite{MB_skin}, 
particles are only moderately localized to one end of the system
as a consequence of the competition between
the asymmetry in hopping
and the Pauli exclusion principle
that prohibits all skin modes located at the same site.
The corresponding value of the many-body IPR [Fig. 2 (b)],
defined as ${\cal IPR}=\sum_{\{n\}} |c_{\{n\}}|^4$,
is indeed smaller than the one for the single-particle skin effect [cf. Fig. 1 (a)];
$|\mu\rangle=\sum_{\{n\}} c_{\{n\}} |\{n\}\rangle$, and $|\{n\}\rangle$
represents a computational basis.
${\cal IPR}$ measures a localization tendency in the many-body Fock-space.
Fig. 2 (b) reveals a different behavior of ${\cal IPR}$ under PBC vs. OBC.
With the increase of disorder strength $W$
${\cal IPR}$ shows under the OBC
a dip in the critical regime: $W\sim W_c$,
before it turns to increase in the MBL regime.
Under the PBC, ${\cal IPR}$ stays $\simeq 0$ until it surges
after a critical disorder strength $W=W_c$.
This is consistent 
with the behavior of the many-body winding number $\nu$ [Fig. 2 (c)],
defined as Eq. \ref{winding} (inset shows typical behavior of $dH(\Phi)$).
Non-zero $\nu$ corresponds to
the appearance of many-body skin modes\cite{MB_topo}.
Note that $\nu$ is not the number of skin mode.

Here, our main focus is on
the excited states of the Hatano-Nelson model,
but it is also interesting to investigate the properties of the ground state; 
we assume the OBC case with real eigen-energies.
Fig.2 (d) shows the change of
charge density-wave order parameter $O_{DW}={1\over L}|\sum_i(-1)^i\langle n_i\rangle|$ 
in the ground state ($W=0$) 
as a function of the inter-particle interaction V.
$O_{DW}$ sharply increases at $V=2$, i.e., many-body skin mode disappears and 
the CDW order emerges.
Recently, Ref. \citeonline{MB_fock} has reported 
that 
a strong interaction prohibits quantum states to thermalize, 
realizing a feature of Fock-space fragmentation.
It is indeed interesting to study how many-body skin modes are suppressed by
the inter-particle interaction $V$.
In the inset of the Fig. 2(d)
we show that
(the change of) the behavior of $O_{DW}$ 
is concomitant with the vanishing of a finite winding number $\nu$.

\begin{figure}[tbh]
\includegraphics[width=150mm]{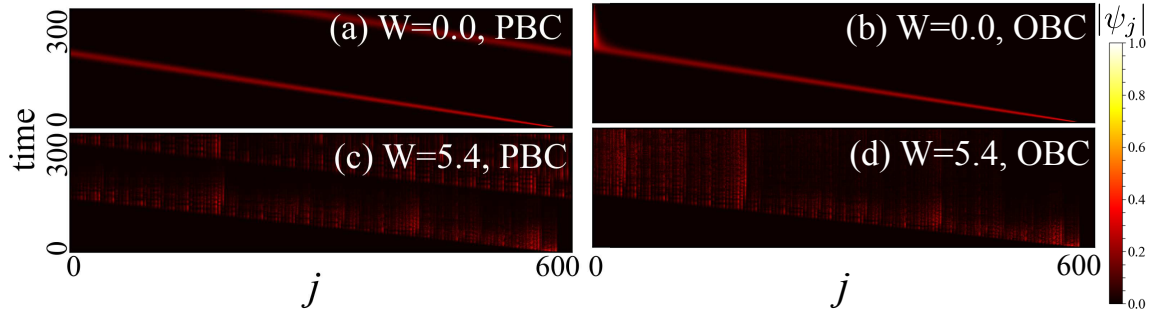}
\caption{Single-particle dynamics. Evolution of a wave packet initially located at $j=j_0=580$. Comparison of different boundary conditions: PBC vs. OBC. (a), (b): clean limit ($W=0$). (c), (d): critical disorder regime ($W=5.4\sim W_c$). 
$M=15$, $\delta t=0.2$; see sec. 3.3 for details.
}
\label{f3}
\end{figure}

\vspace{-0.5cm}
\section{Dynamical properties}

Let us turn our eyes to the dynamics of the system,
and follow how an initial wave packet in our system
evolves in time.
The motivation to study the dynamics in a system described by 
a non-Hermitian Hamiltonian such as the one given in Eq. (\ref{ham})
may be two-fold.
On one hand, 
a system effectively well described by such a non-Hermitian Hamiltonian
emerges dynamically,
i.e., as a non-equilibrium situation, e.g., realized in an open quantum system.
On the other hand,
to study the quantum dynamics 
resulting from a non-Hermitian Hamiltonian 
with the peculiar properties outlined in the previous subsection,
such as the complex spectrum, the wave function showing skin effect, etc.
is {\it per se} of much interest,
resulting in a number of insightful findings.

\subsection{The wave-packet dynamics: single-particle case}

We choose the initial wave packet to be the one localized at site $j=j_0$:
$|\psi (t=0)\rangle=|j_0\rangle$.
At time $t$, the wave packet may evolve as
\begin{equation}
|\psi (t)\rangle
=\sum_j \psi_j(t) |j\rangle
=\sum_{n} c_n e^{-i\epsilon_n t} |n\rangle,
\label{evo}
\end{equation}
where
$|n\rangle$ represents the $n$th single-particle eigenstate of the Hamiltonian (\ref{ham})
with an eigenenergy $\epsilon_n$;
i.e., $H |n\rangle=\epsilon_n |n\rangle$,
while
$c_n =\langle\langle n|\psi (t=0)\rangle$.
Here, 
$\langle\langle n|$ represents the left eigenstate corresponding to the eigenenergy $\epsilon_n$:
$\langle\langle n|H=\epsilon_n\langle\langle n|$
and not $|n\rangle^\dagger$;
$\langle\langle n|\neq |n\rangle^\dagger$.
Note that the left and right eigenstates satisfy the biorthogonal condition, $\langle \langle n|m\rangle=\delta_{n,m}$.
This biorthogonality is another peculiarity of the non-Hermitian system.

We have seen in the last section that
in case of $g\neq 0$
$\epsilon_n$ becomes complex under the PBC
in the regime of weak $W$.
Eq. (\ref{evo}) implies that
those eigenstate with large Im $\epsilon_n$ becomes dominant in the time evolution,
and eventually
$|\psi (t)\rangle$ converges to an eigenstate with the maximal Im $\epsilon_n$.
This makes the time evolution non-unitary.
In the simulation shown below,
we, therefore, 
renormalize
$|\psi (t)\rangle$ as 
$|\psi (t)\rangle\rightarrow|\tilde{\psi}(t)\rangle=
{|\psi (t)\rangle/\sqrt{\langle\psi (t)|\psi (t)\rangle}}$.
In the Hermitian case ($g=0$)
an equal superposition of different eigenstates 
and their quantum interference in the time evolution
leads to
spreading of the wave packet.
Here, 
in case of $g\neq 0$
such quantum interference of the eigen wave function is strongly suppressed.
As a result, the wave-packet dynamics
becomes pseudo-classical,
and 
the probability density $|\psi_j (t)|^2$
obeys effectively the classical diffusion equation,
at least, as in the non-Hermitian case \cite{Orito}
(Fig. 3 (a)).
Instead, the wave packet simply slides 
, reflecting the uni-directionality 
of the model (asymmetry of the hopping);
in this sense this is a natural result.
Under the OBC, $\epsilon_n$ is no longer complex but real. Still, the behavior of the
time evolution is indistinguishable from the case of PBC (Fig. 3 (b)),
until the wave packet reaches the boundary;
in case of OBC,
a mechanism different from the one for PBC
is responsible for the apparently same time evolution.
From the outset,
it is rather natural that a local dynamics of a wave packet, here, in question,
is insensitive to the boundary condition.

Effects of disorder (case of $W\neq 0$) is also very different from the Hermitian case.
In the Hermitian case,
disorder suppresses spreading of the wave packet, since it weakens the quantum interference.
Here, 
in case of $g\neq 0$,
disorder, on the contrary, enhances spreading of the wave packet (Fig. 3 (c)).
In the critical regime $W\simeq W_c$,
a cascade-like enhancement of the wave packet spreading is conspicuous (Fig. 3 (c) and (d)).

\subsection{Many-particle case: density and entanglement dynamics}

Here, we extend the analysis in the previous subsection
on the wave-packet dynamics in the single-particle case
to many-particle systems.
We have studied 
an entanglement entropy $S_{EE}$ defined as $S_{EE}(t)=-\Tr{\rho_R}\ln\rho_R$, where
 $\rho_R=\Tr_L{|\Psi(t)\rangle\langle\Psi(t)|}=\sum_{L,R_1,R_2}\psi_{L,R_1}\psi_{L,R_2}^*|L,R_1\rangle\langle L,R_2|$ is a reduced density matrix ($L$ and $R_i$ define left and right half of system in the real space, respectively) calculated by traceing out the left half of the system.
Here, we consider 
the case of $N=L/2$ particles
in a system of size $L$ (half-filling), and choose the initial state
to be the following domain wall state:
\begin{equation}
|\Psi (t=0)\rangle = |00\cdots 0 11\cdots 1\rangle,
\label{d_wall}
\end{equation}
i.e., the last $L/2$ sites are occupied.

Four panels of Fig. 4 show
some typical examples of the dynamics for the initial state (\ref{d_wall})
for $g=0.5$
and
for different values of disorder strength $W$.
The dynamics of the density $n_j(t)$ is shown in the insets.
In the regime of weak disorder (Fig. 4 (a) and  4 (b))
the system is in the delocalized phase, 
so that the initial domain wall structure tends to 
dissolve into a uniform distribution in the case of PBC (Fig. 4 (a), inset).
Correspondingly,
the entanglement entropy $S_{EE}(t)$ increases as $t$ increases, 
while in the regime of long time scale $t\sim 10^1$,
it turns to decrease, resulting in a non-monotonic evolution of $S_{EE}(t)$.
%
Recall that an initial quantum state is 
generally a superposition of the eigenstates with an individual phase factor,
as time passes by, they tend to behave like a random vector;
thus increasing the entanglement entropy $S_{EE}(t)$\cite{MBL_rev}.
%
However, in this regime, 
the eigenenergies are complex,
so that 
as time evolves,
such a superposition tends to be lost, i.e., $|\Psi(t)\rangle$ converges to 
eigenstate with maximal imaginary eigenenergy (here we name the eigenstate as $|R_{max}\rangle$), and the entanglement entropy
$S_{EE}(t)$ turns to decrease.
Such a convergence process appears as a plateaux of the $S_{EE}(t)$ after certain time.

\begin{figure}[tbh]
\includegraphics[width=150mm]{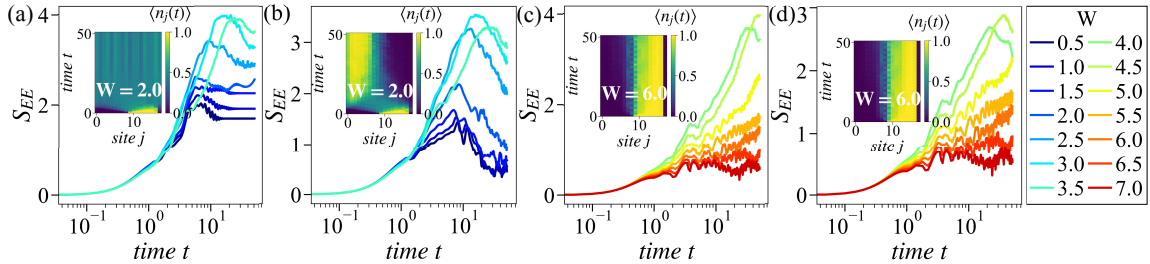}
\caption{Typical entanglement dynamics: (a) delocalized phase with PBC, (b) delocalized phase with OBC, (c) localized phase with PBC, (d) localized phase with OBC. Inset of the each panel show time-evolution of the density pattern. 
$M=25$, $\delta t=0.05-0.2$; see sec. 3.3 for details. 
5 disorder realizations.
$L=18$, $N=8$, and $V=2.0$. 
}
\label{f4}
\end{figure}

%
Under the OBC [Fig. 4 (b)],
$|\Psi(t)\rangle$ does not converge to $|R_{max}\rangle$ by complex spectrum because eigenenergy is real spectrum but realizing dynamics is still non-unitary due to non-orthogonality of the eigenvector ($\langle R_\nu|R_\mu\rangle\neq\delta_{\nu\mu}$), and $|\Psi(t)\rangle$ converges to $|R_\nu\rangle$, i.e., many-body skin mode.
Interestingly, 
entanglement entropy $S_{EE}(t)$ still shows
a non-monotonic evolution 
even though eigenenergies are real in this case.
This occurs 
in a dynamical process of the many-body skin effect;
here,
starting with the initial configuration (\ref{d_wall}),
particles tend to slide to the preferred direction, and
after certain time
they are relocated to their ``right'' positions prescribed by the many-body skin effect.

In the delocalized regime, both under PBC and OBC [Fig. 4 (a) and (b)],
the 
$S_{EE}(t)$
increases with the increase of disorder strength $W$.
%
This is due to the increase of scattering amplitudes by the quasi-periodic/disorder potential (\ref{quasi}),
which scatters
a quasiparticle with crystal momentum $k$ to $k\pm2\pi\theta$.
%
As seen in the single-particle case,
such the scattering gradually makes uni-directional dynamics cascade-like
which leads quasiparticles to be more correlated with each other than free particle case;
therefore, 
$S_{EE}(t)$
increases.
%

In the regime of strong disorder [Fig. 4 (c) and (d)]
the density profile shows a clear localized feature, while the 
$S_{EE}(t)$
shows a 
logarithmic growth
\cite{Abanin}
.
Practically, no dependence on the boundary condition : PBC vs. OBC.

\subsection{Remarks on numerics}
Here,
to deal with a system of larger size than the ones in Ref. \citeonline{Orito},
we have employed the 
Krylov subspace method for non-Hermitian systems\cite{Arnoldi}. 
Since we focus on the time-evolution 
driven by a non-Hermitian Hamiltonian,
we use the Arnoldi method instead of the Lanczos method to generate 
an orthonormal  Krylov space $V_M$ 
from the Krylov space 
$K_M=span(|\Psi(t)\rangle, H|\Psi(t)\rangle,\cdots,H^{M-1}|\Psi(t)\rangle)$.
Using $V_M$,
a unit vector $|e_1\rangle=(1,0,\cdots,0)^T$
and $\tilde{H}=V_M^\dagger HV_M$,
the time evolution of the quantum state is written as
\begin{equation}
|\Psi(t+\delta t)\rangle\sim V_M e^{-i\delta t\tilde{H}}V_M^\dagger|\Psi(t)\rangle=V_M e^{-i\delta t\tilde{H}}|e_1\rangle.
\label{psit2}
\end{equation}
The advantage of this method is that 
the computational complexity of diagonalizing the Hamiltonian 
is reduced to 
that of diagonalizing a smaller $M\times M$ matrix $\tilde{H}$.

In the numerical calculation, we have employed QuSpin\cite{QuSpin} for creating the non-Hermitian matrix such as the one given in Eq. (\ref{ham}).

\section{Concluding remarks}
We have studied the static and dynamical properties of a non-Hermitian system, here, 
taking the Hatano-Nelson type model as a concrete example.
Unlike the static properties sensitive to the boundary condition; e.g.,
complex vs. real spectrum, skin effect, etc.,
the dynamical properties are shown to be, apart from 
the effect of particle reaching the boundary,
not particularly sensitive or insensitive to the boundary condition.
%
We have observed the non-monotonic time evolution 
of the 
$S_{EE}(t)$
and
its enhancement by disorder
both under PBC and OBC.
They have been interpreted as an interplay of disorder and non-Hermicity.
There, such features as skin effect and the imaginary eigenenergy
specific to the non-Hermitian system played a crucial role.
In a future work, we will attempt
a more systematic description of such unique entanglement dynamics
in non-Hermitian systems.

Finally, T.O. is supported by JST SPRING: Grant Number JPMJSP2132, and K.-I.I by JSPS KAKENHI: 21H01005, 20K03788, and 18H03683.


\end{document}